\documentclass[10pt,twocolumn,groupedaddress,aps,prx,amssymb,amsfonts,amsmath,floatfix,nofootinbib,longbibliography]{revtex4-2}
\usepackage{bm}
\usepackage{xcolor}
\usepackage{hyperref}
\usepackage{graphicx}
\usepackage{float}
\usepackage[normalem]{ulem}
\usepackage[english]{babel}
\usepackage{algpseudocode}

\hypersetup{
    colorlinks=true,
    linkcolor=blue,
    citecolor=blue,
    urlcolor=blue,
}

\let\a=\alpha

\def\sign{{\textrm {sign}}}

\def\to{\rightarrow}  

\newcommand{\beq}{\begin{equation}} \newcommand{\eeq}{\end{equation}}

\begin{document}

\title{Dreaming improves memorization in a Hopfield model with bounded synaptic strength}

\author{Enzo Marinari}%
\author{Saverio Rossi}%
\thanks{Corresponding author. Email: saverio.rossi@uniroma1.it}
\author{Francesco Zamponi}%
\affiliation{Dipartimento di Fisica, Sapienza Universit\`a di Roma, Piazzale Aldo Moro 5, 00185 Rome, Italy}

\begin{abstract}
The Hopfield model provides a paradigmatic framework for associative memory. Its classical implementation, based on the Hebbian learning rule, suffers from catastrophic forgetting: when one attempts storing too many patterns, the network fails to retrieve any of them. Yet, the Hebbian rule does not take into account that synaptic strength is bounded. Introducing this biologically plausible modification, known as ``clipping'', eliminates catastrophic forgetting; the model is now able to retrieve the most recently seen memories, eliminating older ones. Yet, its memorization capacity is much reduced with respect to the unclipped case. Here, we investigate the effects of adding a ``dreaming'' phase on the capacity of a clipped Hopfield model. Following a proposal by Hopfield, Feinstein and Palmer, we assume that during the dreaming phase, the model generates random patterns that are then ``unlearned''. We show that while clipping still removes catastrophic forgetting, alternating learning and dreaming phases improves the memorization capacity and makes the search for optimal performance more realistic from an evolutionary perspective.
\end{abstract}

\maketitle

\subsection*{Significance statement}
Associative memory models are widely used to study how neural networks store and retrieve information. In the classical Hopfield model, memories are learned through Hebbian learning, but storing too many patterns leads to catastrophic forgetting: the network abruptly loses the ability to retrieve any memory. Introducing biologically realistic bounds on synaptic strengths removes this instability but strongly reduces storage capacity. Here, we investigate whether spontaneous activity resembling ``dreaming'' can improve memory storage in such constrained networks. We show that alternating phases of learning and dreaming increases the number of retrievable memories while preserving the gradual replacement of older memories by newer ones. These results suggest that internally generated activity may help stabilize memory storage in systems with bounded synapses.

\section{Introduction}

An associative memory is a system capable of retrieving stored patterns starting from noisy or incomplete cues. This ability—often referred to as pattern recovery—is a central feature of both biological and artificial neural networks. The paradigmatic theoretical framework for studying associative memory is the Hopfield model~\cite{Hopfield1982}, in which neurons interact through synaptic couplings shaped by the Hebbian learning rule~\cite{amit1989modeling,Hebb1949}. In this model, memories are encoded as stable attractors of the network dynamics, and retrieval corresponds to the spontaneous convergence of neural activity toward one of these attractors. The performance of associative memory models is commonly quantified by their critical load, defined as the maximum number of patterns that can be stored and reliably retrieved.

A fundamental limitation of the standard Hopfield model emerges when the number of stored patterns becomes too large. Beyond a critical load, the system undergoes a sharp transition known as catastrophic forgetting, in which all previously stored memories abruptly become unstable and cannot be retrieved~\cite{amit1985storing}. In the classical Hebbian model, this critical load scales only linearly with the number of neurons, severely constraining the storage capacity of the network. Over the past decades, substantial efforts have been devoted to overcoming this limitation and to developing mechanisms that increase memory capacity.

One of the earliest and most influential proposals in this direction is the unlearning procedure introduced in Ref.~\cite{Hopfield1983}. In this approach, the network is allowed to explore its state space starting from a random initial condition, thereby generating spurious attractors—stable configurations that do not correspond to meaningful memories—which are subsequently weakened through synaptic updates. This process, often referred to as dreaming, was inspired by the hypothesis that sleep may serve a similar function in biological brains by selectively removing undesirable neural correlations~\cite{crick1983function}. This idea has been explored extensively and unlearning procedures based on weakening of spurious attractors have been shown to increase storage capacity and improve the handling of correlated memories~\cite{van1990increasing,van1992unlearning,wimbauer1994universality,van1997hebbian,Benedetti2022,benedetti2024training}. Another interesting form of unlearning is based on removing neuron correlations evaluated in the high-temperature phase~\cite{nokura1996unlearning,takeuchi2026analysis}. Mixed learning-and-dreaming algorithms, in which phases of memory acquisition alternate with phases of unlearning activity, have also been proposed and studied~\cite{christos1996investigation,fachechi2019dreaming,Serricchio2025}.

From a biological perspective, spontaneous neural activity during sleep and rest is believed to play an important role in memory processing. In particular, neural activity patterns associated with prior experiences are often spontaneously reactivated during sleep, a phenomenon known as replay~\cite{peyrache2009replay,tavoni2015inferred,kaefer2022replay}, which is thought to contribute to memory consolidation by reinforcing and reorganizing synaptic connections. 
It was recently shown that disruption of the replay process is associated with memory decline~\cite{shipley2026disrupted}.
Although the dreaming and unlearning procedures studied in associative memory models are highly idealized, they capture a related qualitative principle: internally generated activity can reshape synaptic couplings and improve memory stability.

A different line of research has focused on modifying the learning rule itself. For example, the pseudo-inverse rule allows for optimal storage of patterns~\cite{personnaz1985information,kanter1987associative}, but relies on global computations, such as matrix inversion, that are difficult to reconcile with biological constraints. Other modern variants of associative memory models achieve dramatically increased capacity by introducing higher-order interactions~\cite{krotov2016dense,demircigil2017model,ramsauer2020hopfield,lucibello2024exponential}. While highly effective from a computational perspective, these mechanisms are also challenging to interpret biologically, as they require interactions beyond pairwise synapses. Consequently, recent work has sought to identify mechanisms that improve memory performance while remaining compatible with biologically plausible constraints~\cite{kafraj2026,krotov2020large}. Among such constraints, two important ones are the bounded nature of synaptic strengths and the locality of synaptic plasticity.

Despite these advances, catastrophic forgetting remains a persistent feature of most associative memory models: when the load becomes too large, the network abruptly loses all previously stored information. This behavior is both biologically implausible and functionally undesirable. In biological systems, memories typically degrade gradually, with newer experiences progressively interfering with older ones rather than erasing them entirely. Understanding how to reproduce such gradual forgetting within simple models is therefore an important open problem.

In this context, clipping procedures—in which synaptic strengths are explicitly bounded—provide a simple and biologically motivated modification of the Hopfield model. By preventing synapses from growing without limit, clipping introduces competition between memories and leads to a progressive degradation of retrieval performance, rather than an abrupt collapse. Although clipping reduces the maximum achievable capacity, it eliminates catastrophic forgetting and produces a more realistic memory dynamics~\cite{Parisi1986,Marinari2019}.

In this work, we investigate how dreaming interacts with synaptic clipping, and whether alternating phases of learning and dreaming can improve memory performance in a biologically constrained associative memory. Our central result is that dreaming substantially enhances the storage capacity of the clipped Hopfield model. More importantly, we find that the capacity exhibits a smooth and robust maximum as a function of the relative amount of learning and dreaming. This is in sharp contrast with the unclipped (standard) case, where optimal performance requires a fine-tuned balance between the two processes~\cite{Serricchio2025}. The presence of a broad optimum suggests that such learning–dreaming cycles could emerge naturally through adaptive or evolutionary mechanisms, without requiring precise parameter tuning. Our results therefore provide a simple and stylized framework in which spontaneous activity, interpreted as dreaming, plays a constructive role in stabilizing memories under biologically realistic constraints.

The remainder of the paper is organized as follows. In Sec.~\ref{sec:methods}, we introduce the model and describe the learning–dreaming algorithm, highlighting the relevant limiting cases, including standard Hebbian learning, dreaming, and alternating learning–dreaming dynamics. In Sec.~\ref{sec:results}, we present our main results and compare the retrieval performance of these different algorithms, both with and without synaptic clipping. Finally, in Sec.~\ref{sec:discussion}, we discuss the implications of our findings and outline possible directions for future work.

\section{Methods}
\label{sec:methods}

Associative memory models store information in the structure of synaptic interactions between neurons. Memories correspond to attractors of the network dynamics, that is, stable configurations toward which neural activity spontaneously converges. The synaptic coupling matrix determines the shape of this attractor landscape: acting on synaptic strengths can either stabilize desired memories or destabilize unwanted configurations. In this work, we study how alternating phases of learning and dreaming affects memory retrieval performance, both in the presence and absence of constraints on synaptic strength (called clipping).
We consider a recurrent Hopfield network of $N$ binary neurons $s_i$ whose activity evolves deterministically under the influence of synaptic inputs from other neurons~\cite{Hopfield1982}. At each time step, every neuron sequentially (in randomly shuffled order) updates its state by aligning with its local field, defined as the weighted sum of the activity of the other neurons as
\begin{equation}
    \label{eq:Hebbdynamics}
    s_i^{(t+1)} = \sign\left( \sum_{j=1}^N J_{ij}s^{(t)}_j \right).
\end{equation}
The dynamics evolves until the system reaches an attractor. The goal is to define a coupling matrix $J$ such that the dynamics in Eq.~\eqref{eq:Hebbdynamics} is able to recover a set of $P$ independent patterns defined as {$\xi_i^{\mu}$}, with $\mu = 1,\dots,P$. One can define different criteria for considering a memory as retrieved. In the following we choose to consider a memory to be successfully recovered if the dynamics starting from it converges to a fixed point that differs from the initial configuration by no more than 2\% of the neurons (see the SI for a different definition). In particular, we are interested in quantifying the amount of recovered memories when the model load $\alpha=P/N$ changes.

Many algorithms to learn 
an appropriate matrix $J_{ij}$ from the patterns have been proposed. Here, the synaptic coupling matrix is shaped through two complementary processes: learning and dreaming. Learning reinforces synaptic correlations associated with stored patterns, thereby stabilizing the corresponding attractors. Dreaming, by contrast, is driven by internally generated activity: the network evolves from random initial conditions toward spontaneously emerging attractors, which are subsequently weakened through synaptic modifications. During the learning part, we select a random memory $\mu_l$ and we act on the coupling matrix to amplify it, according to
\begin{equation}
\label{eq:learningcontribution}
    J_{ij} \leftarrow J_{ij} + \frac{1}{\tau_l \sqrt{N}}\xi^{\mu_l}_{i} \xi_j^{\mu_l}
\end{equation}
in parallel for every couple of neurons $i$ and $j$, with $\tau_l$ the learning rate.  We repeat this operation $L$ times. While usually a factor $1/N$ is used,  here we chose a $1/\sqrt{N}$ normalization, which ensures that the typical magnitude of synaptic couplings remains comparable to the clipping threshold introduced below, independently of system size. As a result, clipping can effectively constrain synaptic growth without requiring parameter rescaling as the network size changes.
In the dreaming part we instead start from a random configuration $\bm{s}^{0}$ of neurons and we let the dynamics reach a fixed point $\bm{s}^{*}$. We then modify the coupling matrix to weaken the fixed point, following
\begin{equation}
    \label{eq:dreamingcontribution}
    J_{ij} \leftarrow J_{ij} - \frac{1}{\tau_d \sqrt{N}}s^*_i s^*_j
\end{equation}
in parallel for every couple of neurons $i$ and $j$, with $\tau_d$ the dreaming rate. This anti-Hebbian update weakens the synaptic structure supporting the spontaneously reached fixed point. Because spurious attractors are visited more frequently during dreaming than strongly reinforced memory attractors, this process preferentially destabilizes unwanted configurations while preserving meaningful memories.
We repeat this dreaming step $D$ times, and we alternate learning and dreaming for $T$ cycles.

In biological neural systems, synaptic strengths are bounded and cannot grow indefinitely. To incorporate this constraint, we introduce a clipping procedure that limits each synaptic coupling to lie within the interval $[-A, A]$. After each learning and dreaming update, synaptic values exceeding this range are set equal to the corresponding bound. This constraint prevents uncontrolled synaptic growth and introduces competition between memories, leading to more realistic memory dynamics. Notice that, when clipping, the normalization step introduced in~\cite{Serricchio2025} is not needed anymore as the norm of the coupling matrix is now bounded. It has been shown~\cite{Parisi1986,Marinari2019} that, for the Hebb rule, the optimum value of clipping is around $A\approx0.4$, so we will use this value for most of the work (some results obtained with different values of $A$ are shown in the SI).

The pseudocode for this very general algorithm is reported in Alg.~\ref{alg:general}.
Each cycle of the algorithm consists of two opposing processes: learning steps deepen attractors associated with stored patterns, while dreaming steps weaken attractors generated spontaneously by the network. Clipping ensures that synaptic strengths remain bounded throughout this process. The interplay between these mechanisms determines the stability of stored memories and the overall retrieval performance.

This general framework encompasses several previously studied learning rules as special cases, allowing direct comparison with known results:
\begin{itemize}
    \item The standard Hebb rule~\cite{Hopfield1982} for $T=1$, $L=P$, $D=0$, and $\tau_l = 1$;
    \item The dreaming algorithm~\cite{Hopfield1983,van1990increasing,van1992unlearning,wimbauer1994universality,van1997hebbian,Benedetti2022} for $T=1$, $L=P$, and $\tau_l = 1$;
    \item The daydreaming algorithm~\cite{Serricchio2025} for $L=D=1$ and $\tau_l = \tau_d = \tau$.
\end{itemize}
{
\renewcommand{\figurename}{Algorithm}
\begin{figure}[t]
\caption{Pseudo-code of the Learning and Dreaming algorithm.}
\label{alg:general}
\begin{algorithmic}[1]
\Require Examples $\{ {\xi_i^{\mu}} \}_{i=1,\dots,N}^{\mu=1,\dots,P}$, parameters $T$, $L$, $D$, $\tau_l$, $\tau_d$, $A$.   
    
\State $J_{ij} \gets \textbf{0}$
\For{$t = 1,\dots,T$}
    \For{$l = 1, \dots, L$} \Comment{Learning part}
        \State $\mu_l = \mathrm{rand}(1,\dots,P)$ 
        \State $J_{ij} \gets J_{ij} + \frac{1}{\tau_l \sqrt{N}} \xi_i^{\mu_l} \xi_j^{\mu_l} $ 
        \If{$|J_{ij}| > A$}
            \State $|J_{ij}| = A$
        \EndIf
    \EndFor
    
    \For{$d = 1,\dots,D$} \Comment{Dreaming part}
        \State $s_{i} \gets \mathrm{Unif}(\{ -1,+1\})$ 
        \While {at least one $s_i$ changed} 
            \State $s_i \gets \mathrm{sign}(\sum_{j} J_{ij}s_j)$ 
        \EndWhile
        \State $J_{ij} \gets J_{ij} - \frac{1}{\tau_d\sqrt{N}} s_i s_j$
        \If{$|J_{ij}| > A$}
            \State $|J_{ij}| = A$
        \EndIf
    \EndFor
\EndFor
\end{algorithmic}
\end{figure}
}
\addtocounter{figure}{-1}

Although alternating learning and dreaming phases have been shown to significantly improve memory performance in the unclipped case~\cite{Serricchio2025}, this mechanism requires a precise balance between the two processes, which may be difficult to achieve in biological systems. Furthermore, its behavior in the presence of biologically motivated constraints such as synaptic clipping remains largely unexplored. Understanding how bounded synapses interact with learning and dreaming dynamics is therefore essential to assess the robustness and biological plausibility of these mechanisms.

\section{Results}
\label{sec:results}

The central question we address is how many memories remain stable after learning and dreaming. To quantify this, we measure the fraction of stored patterns that are stable fixed points of the network dynamics. This quantity provides a direct measure of memory capacity, as it captures the ability of the network to preserve previously learned information.
An alternative approach, commonly used in previous work~\cite{Serricchio2025}, is to characterize retrieval through basin-of-attraction maps, which measure how retrieval depends on the level of noise in the initial condition. While this approach provides detailed information about robustness to perturbations, here we focus instead on the stability of the memories themselves. This choice allows us to directly track how many stored patterns remain retrievable as the load and learning parameters are varied, and to identify the onset of memory degradation.

Operationally, a memory $\bm{\xi}^{\mu}$ is considered stable if, when used as the initial condition $\bm{s}^0 = \bm{\xi}^\mu$, the dynamics converges to a fixed point $\bm{s}^*$ that differs from the original pattern by less than a small tolerance:
\begin{equation}
    \Delta^{\mu} = \frac{1}{2 N}\sum_{i=1}^N |s_i^* - \xi^\mu_i|<\epsilon.
\end{equation}
We then define the recognition rate as the number of stored patterns that remain stable after training divided by the number of neurons,
\begin{equation}
    \label{eq:recrate}
    \rho = \frac{1}{N}\sum_{\mu=1}^P \theta(\epsilon - \Delta^\mu),
\end{equation}
with $\theta(\cdot)$ the Heaviside function. Throughout this work we fix $\epsilon=0.02$. Previous studies of dreaming algorithms~\cite{van1997hebbian,Benedetti2024} focused on perfectly recovered memories ($\epsilon=0$), using stability-based criteria~\cite{krauth1988roles}. Our definition is slightly more permissive, allowing for a small fraction of errors, and is consistent with standard practice in associative memory models. This difference leads to a different quantitative estimate of capacity, although the qualitative behavior remains the same (see SI for a detailed comparison).
We now use this observable to compare the performance of different learning and dreaming algorithms.

\subsection{Hebb rule}

\begin{figure}
\centering
\includegraphics[width=\linewidth]{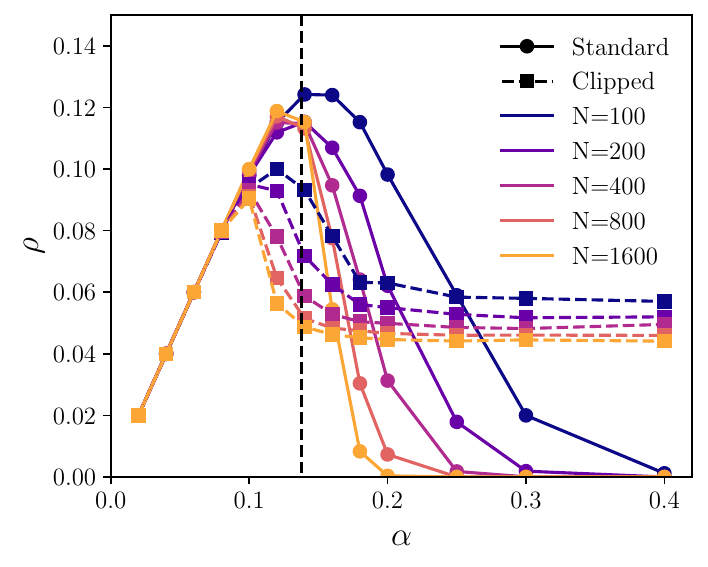}
\caption{Performance of standard (circles) and clipped (squares) Hebb rule as a function of the load. The dashed vertical line highlights the critical load $\a_c$ for the standard Hebb rule when $N\to\infty$. The results are averaged over 50 independent realizations of random memories. 
}
\label{fig:HebbPerformance}
\end{figure}

As a warm-up we consider the case in which the coupling matrix is given by the standard Hebb rule~\cite{Hebb1949}.  This corresponds to fixing $T=1$, $L=P$, $D=0$, $\tau_l=1$ in Alg.~\ref{alg:general}. In this setting the model has a limited storage capacity, since when the load is larger than $\alpha_c \approx 0.138$~\cite{amit1985storing} (far from the theoretical maximum value of $\alpha^{\mathrm {th}}_c = 1$ for symmetric networks~\cite{gardner1988space}) it enters a stage of catastrophic forgetting, where no memory is retrieved. This known result is shown in Fig.~\ref{fig:HebbPerformance}, where the ratio of recovered memories $\rho$ is plotted as a function of the load of the model. In the standard version, as expected, $\rho$ increases linearly with $\alpha$ (notice that $\rho$ is the number of memories recovered divided by $N$, not by $P$) as the model is able to recollect all patterns, until a critical value $\alpha_c \approx 0.138$.  After this value, $\rho$ decreases more and more sharply as the system size increases, confirming that the standard Hebb rule suffers from catastrophic forgetting. 

The effect of clipping has already been studied in this model and it has been shown~\cite{Parisi1986,Marinari2019} that its addition removes the problem of catastrophic forgetting.  The model can thus recover a fraction of memories for any load $\alpha$.
In Fig.~\ref{fig:HebbPerformance}, for the clipped case, we still observe a linear increase of $\rho$ for small loads, but at large loads we see that the model is still able to recover a finite amount of memories. No catastrophic forgetting takes place here.
Interestingly, the large-load capacity in the clipped version of the model is not reached monotonically. 
There is a specific load value for which the model has a maximum capacity, which decreases once the load increases further, and goes asymptotically to a finite, constant value.

\subsection{Dreaming}
\label{subsec:dream}

\begin{figure}
\centering
\includegraphics[width=\linewidth]{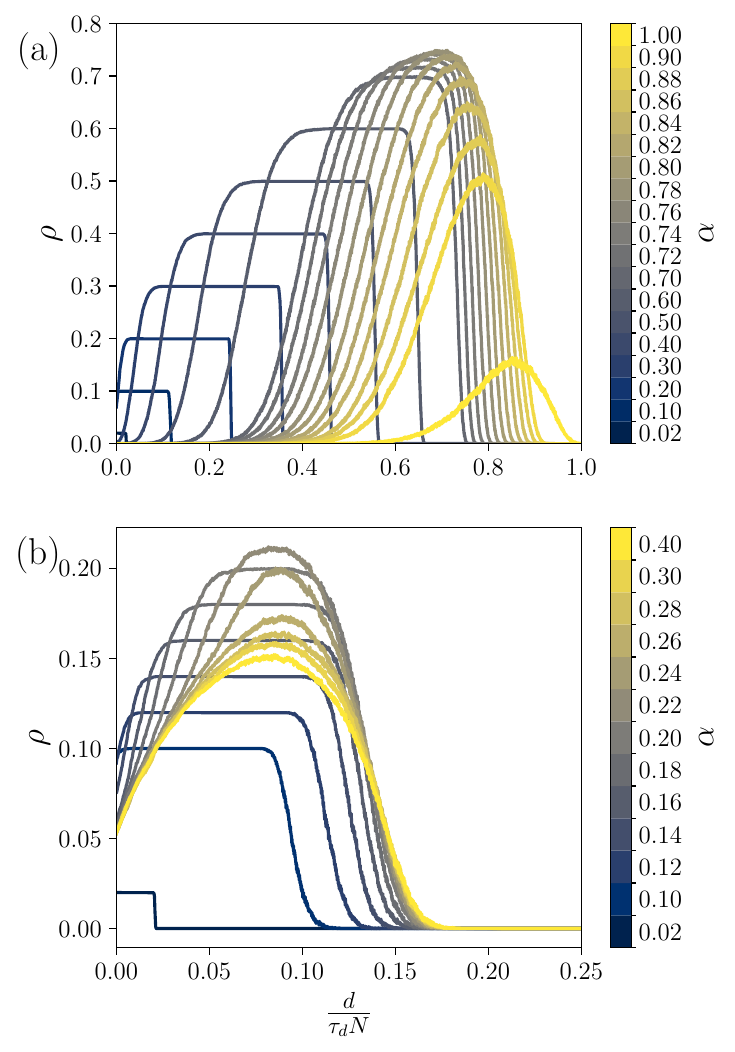}
\caption{Recognition rate $\rho$ as a function of the dreaming steps for the standard (a) and clipped (b) dreaming algorithm, for different values of the load $\alpha$. In the standard case the maximum rate goes to zero for high values of $\alpha$ (catastrophic forgetting) while in the clipped case it goes to a constant, non-zero values, since the system remembers a fixed amount of the most recent patterns. Plots obtained with $N=200$ and averaged over 50 realizations of random memories.
}
\label{fig:DreamingEvolution}
\end{figure}

We now add the dreaming procedure to the picture. This kind of solution to remove spurious memories has already been extensively studied~\cite{Hopfield1983,van1990increasing,van1992unlearning,wimbauer1994universality,Marinari2019,Benedetti2022,Benedetti2024}. We focus on the case in which $T=1$, $L=P$, $\tau_l = 1$ and $D>0$ in Alg.~\ref{alg:general}.  Since we are only performing one cycle here ($T=1$), the absolute values of the rates $\tau_l$ and $\tau_d$ are not relevant.  The only important factor is the ratio between the two, which regulates the contribution of each dreaming and learning step. We choose $\tau_l=1$ and $\tau_d = 100$.

We start by reporting in Fig.~\ref{fig:DreamingEvolution} the evolution of the recognition rate as the dreaming proceeds, for different loads.  In panel (a) we study the standard case, without clipping. When the load is smaller that the Hebb critical one $\alpha < \alpha_c$, the recognition rate starts from $\alpha$, as already without dreaming the model is able to recognize all the memories (see the two darkest curves).  In this case the unlearning procedure does not change the performance initially, but by dreaming too much the model will eventually forget all the memories. On the other hand, when $\alpha > \alpha_c$, the curve starts from zero as the Hebb coupling matrix cannot recover any pattern.  In this case dreaming helps removing spurious memories and increases the model recognition rate, before eventually bringing it back to zero. Dreaming brings an improvement across a wide range of load values, but when $\alpha$ is too large the recognition rate vanishes anyways. The amount of dreaming necessary to reach the optimal performance increases with the load. 

In Fig.~\ref{fig:DreamingEvolution}(b) we report the same analysis for the clipped case, choosing again $A=0.4$, $\tau_l =1$, and $\tau_d=100$. The behavior is different from the one highlighted in panel (a).  First, in this case $\rho$ does not start from zero, even for large loads.  This is due to the fact that when clipping, the catastrophic forgetting is absent and, as a consequence, even before dreaming the model will retain some memories. Moreover, the best performance is achieved over the same range of dreaming steps for a wide range of $\alpha$, differently from what we observe in panel~(a). The number of dreaming steps required to achieve the best possible performance is also much smaller in the clipped case with respect to the standard one. Finally, as for the pure Hebb case, the addition of clipping removes the catastrophic forgetting which occurs for high loads.  As $\alpha$ increases, the recognition rate does not vanish, but instead saturates to a limit curve displaying a pronounced peak. Hence, a proper amount of dreaming brings $\rho$ from about 0.05 to about three times more in the large $\alpha$ limit. Note that while a gain of a factor three might not be impressive for artificial neural networks, being able to store three times more memories might make a big enough difference in a biological context to motivate the adoption of dreaming.

\begin{figure}
\centering
\includegraphics[width=\linewidth]{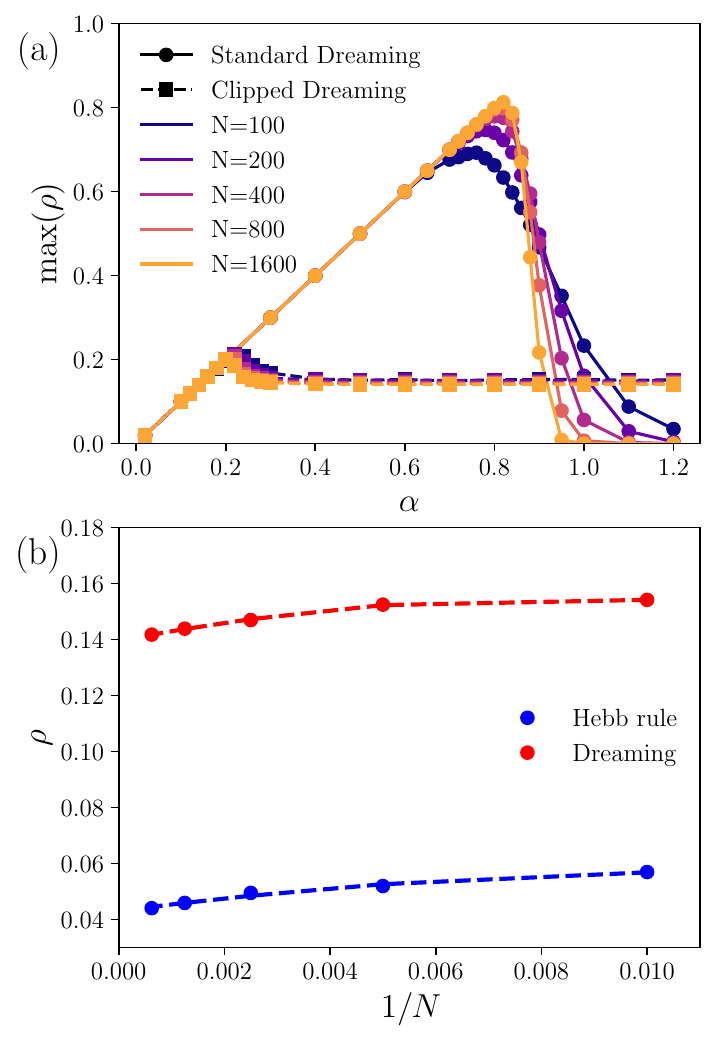}
\caption{(a) Best performance achievable for standard (circles) and clipped (squares) dreaming algorithm as a function of the load. The results are obtained with $\tau_l =1$, $\tau_d=100$, and averaged over 50 independent realizations of random memories. 
(b) Scaling of the best possible $\rho$ with the number of neurons for the Hebb rule (blue) and the dreaming algorithm (red). The dashed lines are fits obtained with $f(N) = \rho_{\infty} + c_1/N + c_2/N^2$, where $\rho_{\infty}= 0.043$, $c_1 = 2.45$, $c_2=-106.65$ for the Hebb rule and $\rho_{\infty}= 0.139$, $c_1 = 3.64$, $c_2=-217.24$ for the dreaming algorithm.
}
\label{fig:DreamingPerformance}
\end{figure}

The performance of these algorithms is summarized in panel (a) of Fig.~\ref{fig:DreamingPerformance}. Here we plot the peak value of the recognition rate as a function of $\alpha$ for different system sizes, similarly to what we did for the Hebb rule in Fig.~\ref{fig:HebbPerformance}. The standard dreaming algorithm gives a $\rho$ that increases linearly with $\alpha$ up to a critical value which is much larger than the Hebb one, around $\alpha \approx 0.8$.  After that, the recognition rate decreases rapidly to zero, faster with increasing system size. The clipped case does not show this decrease and reaches instead a fixed value of $\rho$ that is independent of the load for $\alpha > 0.3$.  Once again, clipping removes the catastrophic forgetting. In this figure we also see that the final value of $\rho$ is not reached in a monotonic way, but there is instead a value of $\alpha$ for which the model is able to store a larger amount of memories. The largest possible $\rho$ obtained in the clipped dreaming algorithm is not as high as the one seen in the unclipped case, but it is still better than the best performance of the clipped Hebb rule. This establishes that, also in the clipped scenario, dreaming improves memorization.

In light of Fig.~\ref{fig:HebbPerformance} and Fig.~\ref{fig:DreamingPerformance}(a) one may ask if the improvement provided by dreaming persists in the thermodynamic limit.  We show in Fig.~\ref{fig:DreamingPerformance}(b) the scaling with the inverse system size of the recognition rate at fixed value of $\alpha=0.4$ for the clipped Hebb rule and the clipped dreaming. Both curves seem to saturate at a finite value. A fit gives $\rho_{N \rightarrow \infty}(\alpha=0.4) \sim 0.043$ and $0.139$ for the clipped Hebb and the clipped dreaming frameworks, respectively.

\subsection{Learning and dreaming}
\label{subsec:daydream}

\begin{figure}
\centering
\includegraphics[width=\linewidth]{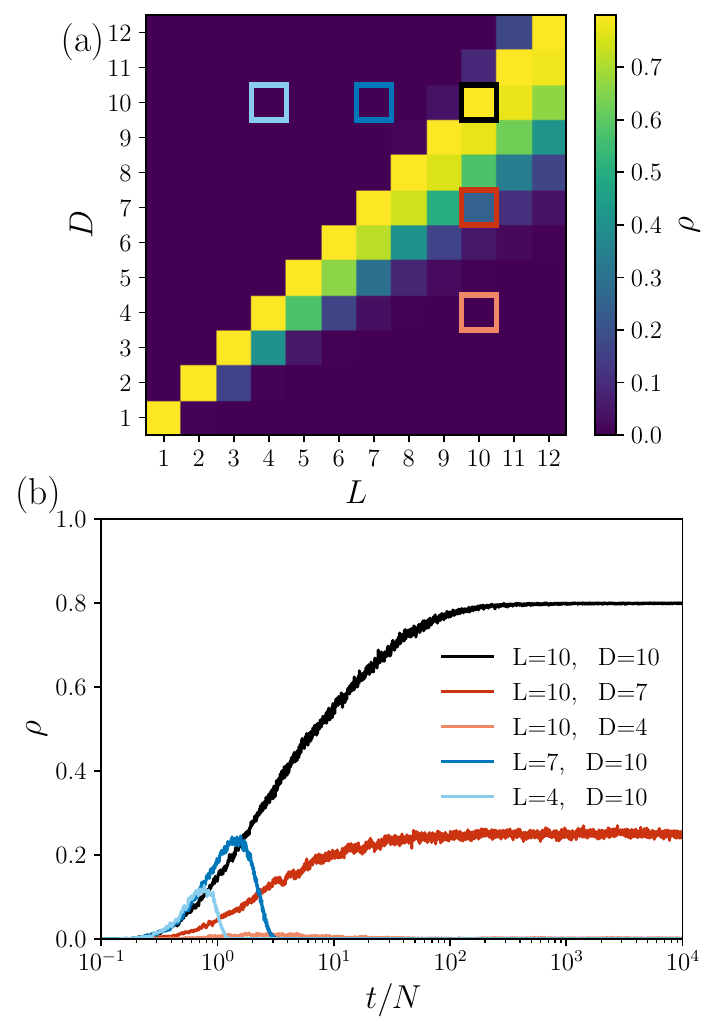}
\caption{Performance of unclipped learning and dreaming algorithm for $N=200$, $\alpha=0.8$, and $\tau_l=\tau_d=10$.  In (a) we report an average over the last 50 points of the $\rho(t)$ curve for a wide range of learning and dreaming steps. Each point is averaged over 50 realizations. In (b) we plot $\rho$ as a function of the epochs for different ratios of $L$ and $D$, highlighted with squares of the same color in the top panel. Each curve is averaged over 10 realizations.
}
\label{fig:LearnandDreamResults}
\end{figure}

We now discuss the most general version of the learning and dreaming algorithm (Alg.~\ref{alg:general}). We consider first the unclipped, catastrophic case, which was already studied in Ref.~\cite{christos1996investigation}.  In order to achieve a faster convergence of the algorithm, we initialize the coupling matrix with the Hebb rule, instead of initializing to $\bm{J} = 0$.  Our Alg.~\ref{alg:general} reduces to the daydreaming procedure recently proposed in Ref.~\cite{Serricchio2025} when $L=D=1$ and $\tau_l = \tau_d = \tau$. We fix the learning and dreaming rates to be equal and we choose in particular $\tau_l = \tau_d = 10$, but we keep the learning and dreaming steps per cycle as tunable parameters.  Our first goal is to investigate how the memorization performance depends on $L$ and $D$. In the top panel of Fig.~\ref{fig:LearnandDreamResults} we show a heatmap of the recognition rate at the end of the dynamics for different choices of the learning and dreaming steps, for $N=200$ and $\alpha=0.8$.  From this figure it is clear that the best performance of the model is achieved when the amount of learning is of the same order as the amount of dreaming.  Obviously, this is sensitive to the choice of the two rates, so that if $\tau_l = n \tau_d$ the best performance is obtained for $D=nL$. Along the diagonal the final value of $\rho$ reached by this procedure does not depend on the value of $L$, but models with larger $L$ will saturate faster in terms of cycles, as they see a larger number of memories for the same time $t$.
Outside the $L=D$ line the model performs poorly. In particular, when $D>L$, the learning dynamics converges to a coupling matrix which is unable to recover memories, as dreaming too much destroys the patterns. On the other hand, for $D<L$ we see that some memories survive for large enough $L$.  Here the dreaming has a positive effect, but not large enough to achieve the optimum.  This is confirmed also in panel (b), where the evolution of the recognition rate is plotted as a function of time during the learning dynamics.  One can clearly see how the two blue lines, in which the dreaming contribution is larger than the learning one, rapidly decrease to zero after some iterations.  The light red curve also goes to zero, due to the fact that the dreaming procedure is not strong enough to remove spurious attractors. Our conclusion is that in this setting the model performs well in the restricted region of the $(D,L)$ space around the diagonal.  As a consequence, a random search for the optimal performance would not converge easily: either the search is already on the line $L=D$, which is singularly optimal, or, if outside of it, it cannot recover memories at all and does not have a clear direction to follow in order to reach the optimum.

\begin{figure}
\centering
\includegraphics[width=\linewidth]{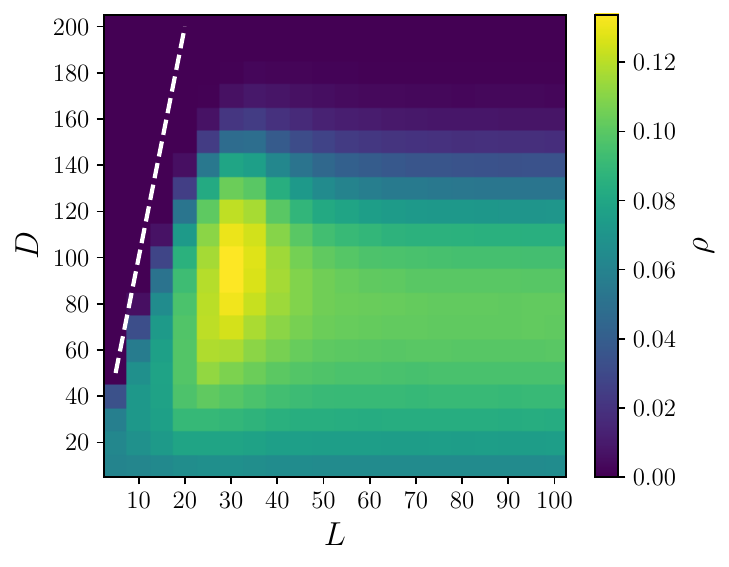}
\caption{Performance of clipped learning and dreaming algorithm for $N=200$, $\alpha=1.2$, $\tau_l=1$, and $\tau_d=10$.  Each square is an average over the last 20 points of the $\rho(t)$ curve for a wide range of learning and dreaming steps, for $\alpha=1.2$. Each point is averaged over 50 realizations.
}
\label{fig:ClippedLearnandDreamResults}
\end{figure}

The situation is different in the clipped case, here initialized in the clipped Hebb matrix. For simplicity we focus only on the high load scenario, fixing $\alpha = 1.2$.
In this case the performance of the model saturates immediately, so again the absolute value of $\tau_l$ and $\tau_d$ are not important, and only their ratio is relevant. We fix $\tau_l =1$ and $\tau_d = 10$. We use the same analysis as before, looking at the model performance in the $(D,L)$ space.  The result, shown in Fig.~\ref{fig:ClippedLearnandDreamResults}, is strikingly different from the one in Fig.~\ref{fig:LearnandDreamResults}. As $\tau_l \neq \tau_d$, the equal contribution line that was before in the diagonal should now be expected to lie along the $D=L \tau_d/\tau_l$ line, which is the dashed white line in the figure. Like in the unclipped case, dreaming too much destroys all the memories. The difference is that there is a much larger part of the $(D,L)$ plane in which the model is able to recollect memories to a certain extent. The best performance is achieved not along the equal contribution line, but instead in a wide region with $L$ between 20 and 40 and $D$ between 50 and 120.  Moreover, even away from this broad optimum, the model still performs very well. For $L \gtrsim 55$, the amount of learning steps does not affect the performance, as the landscape is flat in this direction. However, as a function of $D$, there is an optimal region which can be reached when optimizing.  The optimal recognition value is still modest with respect to the unclipped case, which is in line with what observed in the previous sections. However, the largest value here is still a bit larger than the best one can obtain with a clipped dreaming procedure with the same parameters $\tau_l = 1$ and $\tau_d = 10$ (see SI). Most importantly, within the biologically relevant clipped case, dreaming allows to gain a factor of about three in the number of memories that can be recalled.

\section{Discussion}
\label{sec:discussion}

In this paper we have considered a variant of the Hopfield model where synaptic strength cannot grow indefinitely and is instead clipped~\cite{Parisi1986,Marinari2019}, a modification that enhances the model's biological plausibility. While this clipped model inherently possesses a lower capacity compared to the standard architecture, it does not suffer from undesired catastrophic forgetting~\cite{Parisi1986,Marinari2019}. Our main result demonstrates that the integration of a ``dreaming'' step into the learning process ---during which the system spontaneously generates spurious patterns that are subsequently suppressed from the synaptic matrix--- can substantially improve memorization performance. Furthermore, we observed that optimal performance is achieved across a broad region of the parameter space. This robustness suggests that the optimal performance is highly accessible via stochastic search, making a random or evolutionary process efficient in identifying functional regimes.

This setting differs fundamentally from the daydreaming algorithm~\cite{Serricchio2025}, which, while highly efficient for the unclipped model, requires a precise fine-tuning of parameters to ensure the effective number of learning and dreaming steps are equal. While such fine-tuning is useful for achieving higher capacity, both the absence of synaptic bounds and the requirement for exact parameter balancing appear less biologically plausible than the setting proposed here. In our iterative algorithm, which mixes learning and dreaming steps, clipping plays a dual role: it not only prevents catastrophic forgetting but also expands the viable region in the learning-dreaming space where the model can successfully retrieve a significant fraction of total memories.  From an evolutionary standpoint, it is more reasonable to posit that biological systems evolve by searching for an optimum surrounded by a wide basin of viable solutions, rather than by locating isolated, unstable solutions in a vast configuration space. In this sense, the clipped version of the model offers a more robust and biologically grounded --while, of course, still highly stylized-- framework for memory. 

As a follow-up, it would be of great interest to investigate the system's behavior when memories are not independent but correlated. For example they could be generated via the random features model~\cite{goldt2020modeling}. Such an inquiry would test the model against more realistic data distributions and could as well prove useful for applications in machine learning.

\acknowledgements{We thank Federico Ricci-Tersenghi and Matteo Negri for providing useful feedback.
This research has been supported by first FIS (Italian Science Fund) 2021 funding scheme (FIS783 - SMaC - Statistical Mechanics and Complexity) from MUR, Italian Ministry of University and Research and from the PRIN funding scheme (2022LMHTET - Complexity, disorder and fluctuations: spin glass physics and beyond) from MUR, Italian Ministry of University and Research.}

\bibliography{references}

@article{Hopfield1982,
author = {J J Hopfield },
title = {Neural networks and physical systems with emergent collective computational abilities.},
journal = {Proceedings of the National Academy of Sciences},
volume = {79},
number = {8},
pages = {2554-2558},
year = {1982},
doi = {10.1073/pnas.79.8.2554},
URL = {https://www.pnas.org/doi/abs/10.1073/pnas.79.8.2554}}

@article{shipley2026disrupted,
  title={Disrupted hippocampal replay is associated with reduced offline map stabilization in an Alzheimer’s mouse model},
  author={Shipley, Sarah and Abrate, Marco P and Hayman, Robin and Chan, Dennis and Barry, Caswell},
  journal={Current Biology},
  year={2026},
  volume={36},
  pages={859-871.e5},
  publisher={Elsevier}
}

@article{benedetti2024training,
  title={Training neural networks with structured noise improves classification and generalization},
  author={Benedetti, Marco and Ventura, Enrico},
  journal={Journal of Physics A: Mathematical and Theoretical},
  volume={57},
  number={41},
  pages={415001},
  year={2024},
  publisher={IOP Publishing}
}

@article{lucibello2024exponential,
  title={Exponential capacity of dense associative memories},
  author={Lucibello, Carlo and M{\'e}zard, Marc},
  journal={Physical Review Letters},
  volume={132},
  number={7},
  pages={077301},
  year={2024},
  publisher={APS}
}

@article{Benedetti2024,
doi = {10.1088/1742-5468/ad138e},
year = {2024},
month = {jan},
publisher = {IOP Publishing},
volume = {2024},
number = {1},
pages = {013302},
author = {Benedetti, Marco and Carillo, Louis and Marinari, Enzo and Mézard, Marc},
title = {Eigenvector dreaming},
journal = {Journal of Statistical Mechanics: Theory and Experiment}
}

@article{Benedetti2022,
    author = {Benedetti, Marco and Ventura, Enrico and Marinari, Enzo and Ruocco, Giancarlo and Zamponi, Francesco},
    title = {Supervised perceptron learning vs unsupervised Hebbian unlearning: Approaching optimal memory retrieval in {H}opfield-like networks},
    journal = {The Journal of Chemical Physics},
    volume = {156},
    number = {10},
    pages = {104107},
    year = {2022},
    month = {03},
    issn = {0021-9606},
    doi = {10.1063/5.0084219}
}

@article{Hopfield1983,
	title = {‘{Unlearning}’ has a stabilizing effect in collective memories},
	volume = {304},
	copyright = {http://www.springer.com/tdm},
	issn = {0028-0836, 1476-4687},
	url = {https://www.nature.com/articles/304158a0},
	doi = {10.1038/304158a0},
	number = {5922},
	urldate = {2025-03-11},
	journal = {Nature},
	author = {Hopfield, J. J. and Feinstein, D. I. and Palmer, R. G.},
	month = jul,
	year = {1983},
	pages = {158--159}
}

@article{Parisi1986,
doi = {10.1088/0305-4470/19/10/011},
year = {1986},
month = {jul},
publisher = {},
volume = {19},
number = {10},
pages = {L617},
author = {G Parisi},
title = {A memory which forgets},
journal = {Journal of Physics A: Mathematical and General}
}

@article{Marinari2019,
  author={Marinari, Enzo},
  journal={Neural Computation}, 
  title={Forgetting Memories and Their Attractiveness}, 
  year={2019},
  volume={31},
  number={3},
  pages={503-516},
  doi={10.1162/neco_a_01162}}

@article{Serricchio2025,
title = {Daydreaming {H}opfield Networks and their surprising effectiveness on correlated data},
journal = {Neural Networks},
volume = {186},
pages = {107216},
year = {2025},
issn = {0893-6080},
doi = {https://doi.org/10.1016/j.neunet.2025.107216},
url = {https://www.sciencedirect.com/science/article/pii/S0893608025000954},
author = {Ludovica Serricchio and Dario Bocchi and Claudio Chilin and Raffaele Marino and Matteo Negri and Chiara Cammarota and Federico Ricci-Tersenghi}
}

@misc{kafraj2026,
      title={A Biologically Plausible Dense Associative Memory with Exponential Capacity}, 
      author={Mohadeseh Shafiei Kafraj and Dmitry Krotov and Peter E. Latham},
      year={2026},
      eprint={2601.00984},
      archivePrefix={arXiv} 
}

@book{Hebb1949,
  author    = {Hebb, Donald O.},
  title     = {The Organization of Behavior: A Neuropsychological Theory},
  year      = {1949},
  publisher = {Wiley and Sons},
  address   = {New York},
  isbn      = {978-0-471-36727-7}
}

@article{krotov2016dense,
  title={Dense associative memory for pattern recognition},
  author={Krotov, Dmitry and Hopfield, John J},
  journal={Advances in neural information processing systems},
  volume={29},
  year={2016}
}

@article{demircigil2017model,
  title={On a model of associative memory with huge storage capacity},
  author={Demircigil, Mete and Heusel, Judith and L{\"o}we, Matthias and Upgang, Sven and Vermet, Franck},
  journal={Journal of Statistical Physics},
  volume={168},
  number={2},
  pages={288--299},
  year={2017},
  publisher={Springer},
  doi={10.1007/s10955-017-1806-y}
}

@misc{krotov2020large,
  title={Large associative memory problem in neurobiology and machine learning},
  author={Krotov, Dmitry and Hopfield, John},
  year={2020},
  eprint={2008.06996},
  archivePrefix={arXiv} 
}

@article{gardner1988space,
  title={The space of interactions in neural network models},
  author={Gardner, Elizabeth},
  journal={Journal of physics A: Mathematical and general},
  volume={21},
  number={1},
  pages={257},
  year={1988},
  publisher={IOP Publishing},
  doi = {10.1088/0305-4470/21/1/030}
}

@article{fachechi2019dreaming,
  title={Dreaming neural networks: forgetting spurious memories and reinforcing pure ones},
  author={Fachechi, Alberto and Agliari, Elena and Barra, Adriano},
  journal={Neural Networks},
  volume={112},
  pages={24--40},
  year={2019},
  publisher={Elsevier},
  doi={10.1016/j.neunet.2019.01.006}
}

@article{christos1996investigation,
  title={Investigation of the {Crick}-{Mitchison} reverse-learning dream sleep hypothesis in a dynamical setting},
  author={Christos, George A},
  journal={Neural Networks},
  volume={9},
  number={3},
  pages={427--434},
  year={1996},
  publisher={Elsevier},
  doi={10.1016/0893-6080(95)00072-0}
}

@article{crick1983function,
  title={The function of dream sleep},
  author={Crick, Francis and Mitchison, Graeme},
  journal={Nature},
  volume={304},
  number={5922},
  pages={111--114},
  year={1983},
  publisher={Nature Publishing Group UK London},
  doi = {10.1038/304111a0}
}

@article{goldt2020modeling,
  title={Modeling the influence of data structure on learning in neural networks: The hidden manifold model},
  author={Goldt, Sebastian and M{\'e}zard, Marc and Krzakala, Florent and Zdeborov{\'a}, Lenka},
  journal={Physical Review X},
  volume={10},
  number={4},
  pages={041044},
  year={2020},
  publisher={APS},
  doi={10.1103/PhysRevX.10.041044}
}

@misc{ramsauer2020hopfield,
  title={{H}opfield networks is all you need},
  author={Ramsauer, Hubert and Sch{\"a}fl, Bernhard and Lehner, Johannes and Seidl, Philipp and Widrich, Michael and Adler, Thomas and Gruber, Lukas and Holzleitner, Markus and Pavlovi{\'c}, Milena and Sandve, Geir Kjetil and others},
  year={2020},
  eprint={2008.02217},
  archivePrefix={arXiv} 
}

@inproceedings{van1992unlearning,
  title={Unlearning and its relevance to {REM} sleep: Decorrelating correlated data},
  author={Van Hemmen, JL and Klemmer, Nikolaus},
  booktitle={Neural network dynamics: proceedings of the workshop on complex dynamics in neural networks, June 17--21 1991 at IIASS, vietri, Italy},
  pages={30--43},
  year={1992},
  organization={Springer}
}

@article{kanter1987associative,
  title={Associative recall of memory without errors},
  author={Kanter, Ido and Sompolinsky, Haim},
  journal={Physical Review A},
  volume={35},
  number={1},
  pages={380},
  year={1987},
  publisher={APS},
  doi={10.1103/PhysRevA.35.380}
}

@article{personnaz1985information,
  title={Information storage and retrieval in spin-glass like neural networks},
  author={Personnaz, L and Guyon, I and Dreyfus, G},
  journal={Journal de Physique Lettres},
  volume={46},
  number={8},
  pages={359--365},
  year={1985},
  publisher={Les Editions de Physique},
  doi={10.1051/jphyslet:01985004608035900}
}

@article{wimbauer1994universality,
  title={Universality of unlearning},
  author={Wimbauer, Stefan and Klemmer, Nikolaus and van Hemmen, J Leo},
  journal={Neural Networks},
  volume={7},
  number={2},
  pages={261--270},
  year={1994},
  publisher={Elsevier},
  doi = {10.1016/0893-6080(94)90020-5}
}

@article{van1990increasing,
  title={Increasing the efficiency of a neural network through unlearning},
  author={Van Hemmen, JL and Ioffe, LB and K{\"u}hn, R and Vaas, M},
  journal={Physica A: Statistical Mechanics and its Applications},
  volume={163},
  number={1},
  pages={386--392},
  year={1990},
  publisher={Elsevier},
  doi={10.1016/0378-4371(90)90345-S}
}

@book{amit1989modeling,
  title={Modeling brain function: The world of attractor neural networks},
  author={Amit, Daniel J.},
  year={1989},
  publisher={Cambridge university press}
}

@article{amit1985storing,
  title={Storing infinite numbers of patterns in a spin-glass model of neural networks},
  author={Amit, Daniel J and Gutfreund, Hanoch and Sompolinsky, Haim},
  journal={Physical review letters},
  volume={55},
  number={14},
  pages={1530},
  year={1985},
  publisher={APS}
}

@article{van1997hebbian,
  title={Hebbian learning, its correlation catastrophe, and unlearning},
  author={Van Hemmen, JL},
  journal={Network: Computation in Neural Systems},
  volume={8},
  number={3},
  pages={V1},
  year={1997},
  publisher={IOP Publishing},
  doi={10.1088/0954-898X/8/3/001}
}

@article{nokura1996unlearning,
  title={Unlearning in the paramagnetic phase of neural network models},
  author={Nokura, Kazuo},
  journal={Journal of Physics A: Mathematical and General},
  volume={29},
  number={14},
  pages={3871--3891},
  year={1996},
  doi={10.1088/0305-4470/29/14/013}
}

@misc{takeuchi2026analysis,
  title={Analysis of the {H}opfield Model Incorporating the Effects of Unlearning},
  author={Takeuchi, Shuta and Takahashi, Takashi and Kabashima, Yoshiyuki},
  year={2026},
  eprint={2602.08428},
  archivePrefix={arXiv} 
}

@article{krauth1988roles,
  title={The roles of stability and symmetry in the dynamics of neural networks},
  author={Krauth, Werner and Nadal, J-P and Mezard, Marc},
  journal={Journal of Physics A: Mathematical and General},
  volume={21},
  number={13},
  pages={2995--3011},
  year={1988},
  doi={10.1088/0305-4470/21/13/022}
}

@article{kaefer2022replay,
  title={Replay, the default mode network and the cascaded memory systems model},
  author={Kaefer, Karola and Stella, Federico and McNaughton, Bruce L and Battaglia, Francesco P},
  journal={Nature Reviews Neuroscience},
  volume={23},
  number={10},
  pages={628--640},
  year={2022},
  publisher={Nature Publishing Group UK London},
  doi={10.1038/s41583-022-00620-6}
}

@article{peyrache2009replay,
  title={Replay of rule-learning related neural patterns in the prefrontal cortex during sleep},
  author={Peyrache, Adrien and Khamassi, Mehdi and Benchenane, Karim and Wiener, Sidney I and Battaglia, Francesco P},
  journal={Nature neuroscience},
  volume={12},
  number={7},
  pages={919--926},
  year={2009},
  publisher={Nature Publishing Group US New York},
  doi={10.1038/nn.2337}
}

@article{tavoni2015inferred,
  title={Inferred model of the prefrontal cortex activity unveils cell assemblies and memory replay},
  author={Tavoni, Gaia and Ferrari, Ulisse and Battaglia, Francesco P and Cocco, Simona and Monasson, R{\'e}mi},
  journal={bioRxiv},
  pages={028316},
  year={2015},
  publisher={Cold Spring Harbor Laboratory},
  doi={10.1101/028316}
}

\clearpage

\onecolumngrid

\begin{center}
    \vspace*{1cm}
    {\large \bf Supplemental Material for \\ 
    ``Dreaming improves memorization in a Hopfield model with bounded synaptic strength''}
    \vspace{0.5cm}
\end{center}


\renewcommand{\thesection}{S\arabic{section}}
\renewcommand{\thefigure}{S\arabic{figure}}
\renewcommand{\thetable}{S\arabic{table}}
\setcounter{section}{0}
\setcounter{figure}{0}
\setcounter{table}{0}
\setcounter{page}{1}

\section*{Critical load for the dreaming algorithm}

In Fig.~3 of the main text, we showed the performance of the dreaming algorithm as a function of the load $\alpha$. For the case without clipping, previous works reported a critical load $\alpha^{\rm pr}_c \approx 0.59$~\cite{van1997hebbian,Benedetti2022}. However, the results shown in the main text suggest a larger critical value. This apparent discrepancy originates from the different definitions used to quantify memory recovery.

In particular, in~\cite{Benedetti2024} the critical load was estimated by analyzing the evolution of the minimal stability during the dreaming procedure. The stability of neuron $i$ in memory $\mu$ is defined as~\cite{krauth1988roles}
\begin{equation}
\label{eq:Stability}
S^{\mu}_i= \frac{\xi^{\mu}_i}{\sigma_i} \sum_{j=1}^N J_{ij}\xi^{\mu}_j,
\end{equation}
with
\begin{equation}
\label{eq:normstability}
\sigma_i= \sqrt{\frac{1}{N}\sum_{j=1}^N J^2_{ij}},
\end{equation}
and the quantity of interest is the minimal stability across all neurons and memories, $S^{\rm min} = \min(S^{\mu}_i)$. During dreaming, $S^{\rm min}$ typically reaches a maximum at an optimal number of dreaming steps $D_{\rm top}$, with corresponding value $S^{\rm min}_{\rm top}$. The condition $S^{\rm min}_{\rm top}>0$ ensures that all neurons in all memories are stable, while $S^{\rm min}_{\rm top}<0$ indicates that at least one neuron in one memory is unstable. This criterion therefore corresponds to requiring perfect recovery of all stored memories.

To make contact with this definition, it is natural to introduce a recognition rate that counts only perfectly recovered memories. We therefore define
\begin{equation}
\label{eq:recrate_pr}
\rho_{\rm pr} = \frac{1}{N}\sum_{\mu=1}^P \theta(\Delta^\mu),
\end{equation}
which measures the fraction of memories that are \textit{perfectly recovered}. This definition corresponds to choosing $\epsilon=0$ in Eq.~(5) of the main text.

\begin{figure}
\centering
\includegraphics[width=0.5\linewidth]{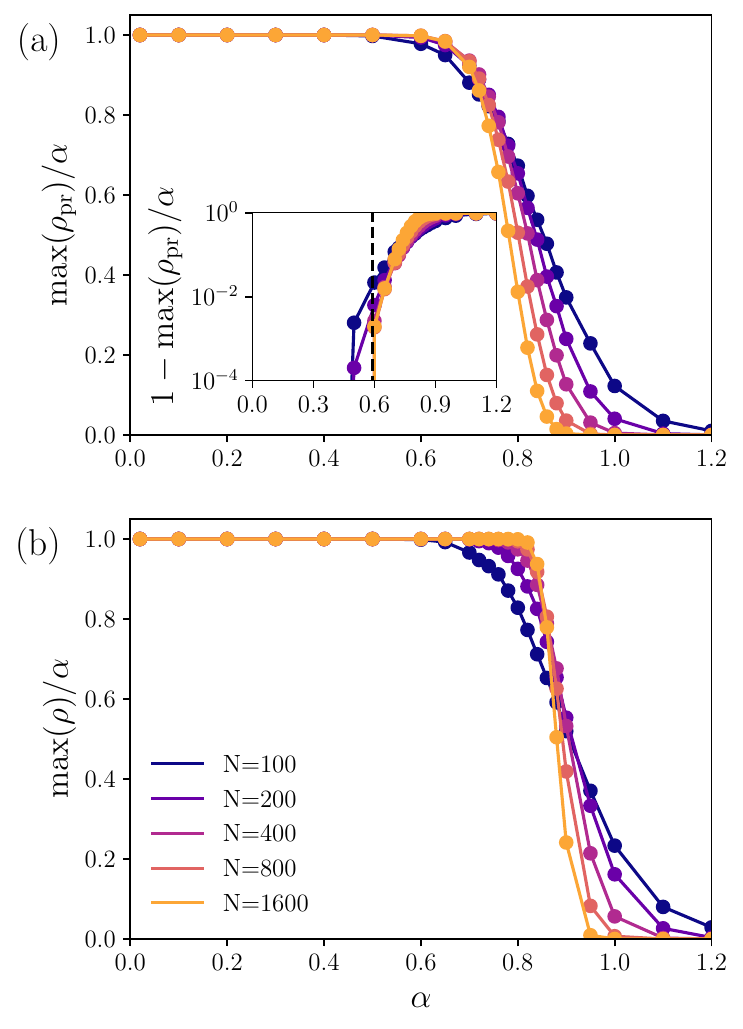}
\caption{Comparison of two ways of measuring the capacity of the network. In (a) the maximum recognition rate $\max(\rho_{\rm pr})$, in which the fraction of memories recovered perfectly, is plotted. The inset shows that, for large enough system sizes, $\max(\rho_{\rm pr})/\alpha<1$ as soon as $\alpha > 0.6$, as found in~\cite{Benedetti2024}. In (b) we plot $\max(\rho)$ as defined in the main text. Here $\tau_d=100$ and each point is averaged over 50 realizations of random memories.
}
\label{fig:PerfectRecoveryPerformance}
\end{figure}

In Fig.~\ref{fig:PerfectRecoveryPerformance}, we compare this recognition rate $\rho_{\rm pr}$ with the recognition rate $\rho$ used in the main text, which allows for a small fraction of errors. Specifically, panel (a) shows the maximum value of $\rho_{\rm pr}$ reached during dreaming, while panel (b) shows the corresponding maximum value of $\rho$. For clarity, we plot the recognition rates divided by the load $\alpha$, so that the curves saturate at one when all memories are recovered and vanish when none are recovered.

As expected, the recognition rate based on perfect recovery, $\rho_{\rm pr}$, decreases at a lower load than $\rho$, since the latter tolerates a small fraction of errors in each memory. To recover the critical load reported in~\cite{Benedetti2024}, one should therefore consider the value of $\alpha$ at which $\max(\rho_{\rm pr})/\alpha<1$, which corresponds to the condition $S^{\rm min}_{\rm top}<0$.

This is confirmed in the inset of Fig.~\ref{fig:PerfectRecoveryPerformance}(a), where we plot $1-\max(\rho_{\rm pr})/\alpha$. For sufficiently large system sizes, this quantity becomes finite only for $\alpha\ge0.6$, in agreement with the previously reported value $\alpha^{\rm pr}_c \approx 0.59$.

In the main text, as well as in the remainder of this Supplementary Material, we adopt the more permissive definition of recognition rate $\rho$, which allows for a small recovery error and is consistent with the standard convention used when analyzing the Hebbian learning rule.

\begin{figure}
\centering
\includegraphics[width=0.5\linewidth]{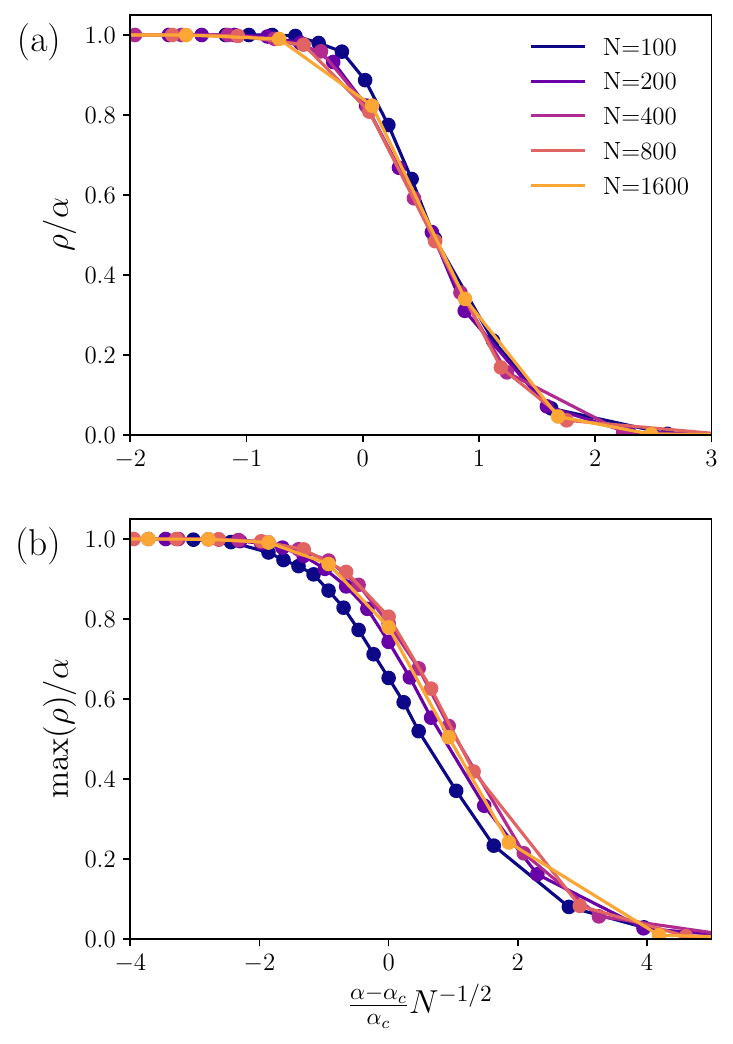}
\caption{Finite-size scaling of the recognition rate around the critical load $\alpha_c$ for the standard Hebb rule (a) and for the dreaming algorithm (b).
}
\label{fig:Scaling}
\end{figure}

Finding a precise value of the critical load is beyond the scope of the present work. Nevertheless, in order to obtain a rough estimate, we performed a finite-size scaling analysis in the vicinity of the transition.
Specifically, in Fig.~\ref{fig:Scaling} we plot the curves of $\rho/\alpha$ as a function of the rescaled variable $N^{-1/2}(\alpha-\alpha_c)/\alpha_c$, which is expected to control finite-size effects close to the critical point. If the chosen value of $\alpha_c$ correctly identifies the transition, the curves corresponding to different system sizes should approximately collapse onto a single master curve.
In panel (a), this analysis is performed for the standard Hebb rule without clipping, using the known critical value $\alpha_c = 0.138$. As expected, the data collapse is very good, confirming the validity of the scaling procedure.
In panel (b), we apply the same analysis to the dreaming procedure without clipping, using the maximum recognition rate reached during the dreaming dynamics. In this case, $\alpha_c$ is treated as an adjustable parameter and chosen to produce the best qualitative collapse of the curves. This procedure yields an estimated critical load $\alpha_c \approx 0.86$.
We emphasize that this value should be regarded only as a rough estimate. The purpose of this analysis is not to determine $\alpha_c$ precisely, but rather to show that the observed transition is compatible with a well-defined critical load.

\section*{Role of $\tau_d$ on dreaming performance}

\begin{figure}[t]
\centering
\includegraphics[width=0.5\linewidth]{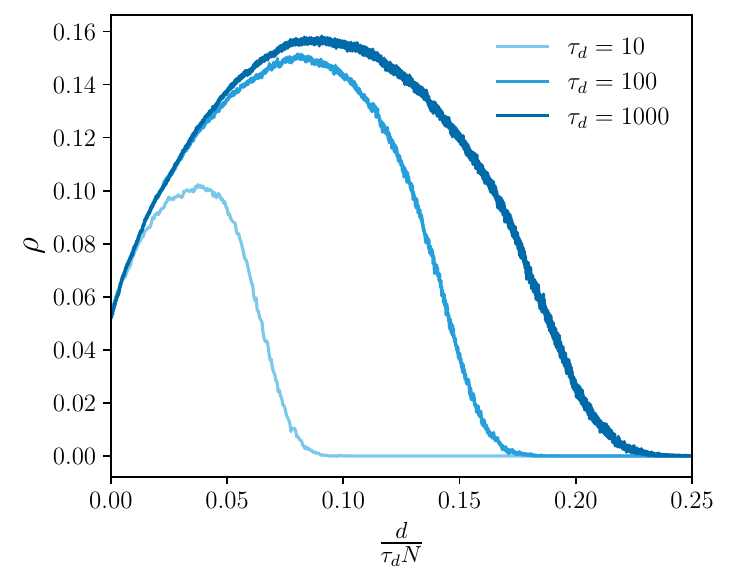}
\caption{Recognition rate $\rho$ as a function of the dreaming steps for the clipped dreaming algorithm, for different values of the dreaming rate $\tau_d$. Here $N=200$, $\alpha=1.2$, and each curve is averaged over 50 realizations of random memories.
}
\label{fig:TauDeffect}
\end{figure}

The results for the dreaming algorithm in the main text are reported for $\tau_d = 100$. 
In Fig.~\ref{fig:TauDeffect} we show what happens for different values of $\tau_d$ for $N=200$.
As already discussed in previous work on  unlearning~\cite{van1990increasing,van1992unlearning,van1997hebbian,Marinari2019,Benedetti2024} a larger value of $\tau_d$ allows to recover a larger fraction of memories.
This is confirmed by the $\rho$ curves in Fig.~\ref{fig:TauDeffect}, in which one can see that not only the maximum value of $\rho$ increases with $\tau_d$, but also the range of values of dreaming steps which give a good amount of recovered memories increase. 
Notice that the improvement of performance with the dreaming rate comes at the cost of a computationally more expensive procedure.
It has been shown that this improvement is limited and that the limit for $\tau_d \rightarrow \infty$ exists, where the performance saturates~\cite{van1997hebbian}. 
For this reason we chose to fix $\tau_d = 100$, which is a good compromise between performance and computational speed.

\section*{Results for different values of $A$}

\begin{figure*}
\centering
\includegraphics[width=0.8\linewidth]{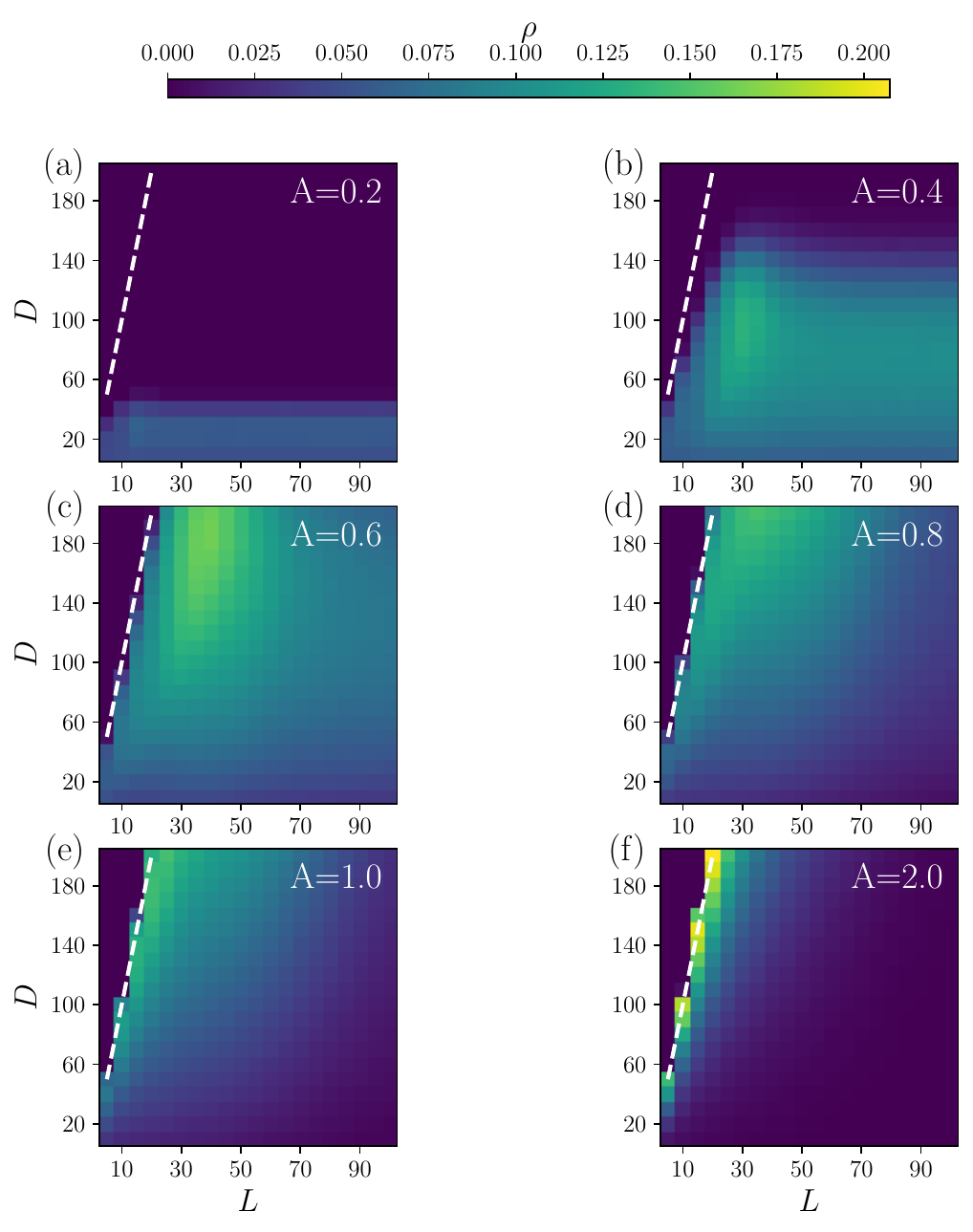}
\caption{Average over the last 20 points of the $\rho(t)$ curve, for different values of $D$ and $L$ and for clipping thresholds $A=0.2$ (a), $A=0.4$ (b), $A=0.6$ (c), $A=0.8$ (d), $A=1.0$ (e), and $A=2.0$ (f). Obtained with $N=200$, $\alpha = 1.2$, $\tau_l=1$, $\tau_d=10$, and averaged over 50 instances of random memories.
}
\label{fig:DiffAValues}
\end{figure*}

Previous studies of the clipped Hopfield model have shown that, for a learning rate $\tau_l = 1$, the clipping threshold that maximizes performance is approximately $A \approx 0.4$. We verified that this value remains optimal also in the presence of the dreaming procedure. For this reason, all clipped simulations reported in the main text were performed using $A = 0.4$.

In Fig.~5 of the main text, we showed the final recognition rate of the alternating learning and dreaming algorithm as a function of the number of learning steps $L$ and dreaming steps $D$. In the absence of clipping, good performance is obtained only when the contributions of the two phases are balanced, i.e., when $L/\tau_l \approx D/\tau_d$. This condition reflects the fact that dreaming alone tends to degrade stored memories, and must therefore be compensated by sufficient learning. In contrast, when clipping is present, a broad region of the $(D,L)$ parameter space yields good recognition performance, indicating that clipping stabilizes the synaptic dynamics and allows memories to persist even when the two phases are not perfectly balanced.
Since the unclipped case corresponds to the limit $A \rightarrow \infty$, it is instructive to examine how this behavior emerges as the clipping threshold is varied. This is shown in Fig.~\ref{fig:DiffAValues}, where we report the final recognition rate for different values of $A$, for $N=200$, $\alpha=1.2$, $\tau_l=1$, and $\tau_d=10$, as in the main text.

For very small values of $A$ (panel (a)), clipping is so strong that synaptic updates are severely constrained. In this regime, learning becomes inefficient, and the dreaming phase progressively degrades the stored memories, resulting in poor performance across most of the parameter space.
For intermediate values of $A$, clipping effectively stabilizes the synaptic matrix while still allowing learning to occur. In this regime, good performance is observed over a wide region of the $(D,L)$ plane, including configurations in which the learning phase dominates ($L/\tau_l > D/\tau_d$). Clipping thus enables the system to tolerate an imbalance between learning and dreaming, preventing the rapid degradation of stored memories.
As $A$ increases further, the effect of clipping becomes progressively weaker, and the system approaches the unclipped regime. Correspondingly, the region of good performance shrinks and becomes increasingly confined to the balanced condition $L/\tau_l \approx D/\tau_d$. In the limit $A \rightarrow \infty$, clipping no longer constrains synaptic growth, and only a precise balance between learning and dreaming allows the network to preserve stored memories.

Overall, these results show that clipping broadens the range of parameters for which alternating learning and dreaming can successfully maintain memory retrieval, by stabilizing synaptic dynamics while still permitting effective learning.

\end{document}